\documentclass{pnastwo}

\usepackage[pdftex]{graphicx}
\usepackage{amsthm,amsmath,amsfonts,amssymb,mathrsfs}
\usepackage{bm,bbm}

\newcommand{\Exp}[1]{\mathbb{E}\left[#1\right]}

\url{www.pnas.org/cgi/doi/10.1073/pnas.0709640104}
\copyrightyear{2008}
\issuedate{Issue Date}
\volume{Volume}
\issuenumber{Issue Number}
\begin{document}

\title{A phase transition in the first passage of a Brownian process through a fluctuating boundary: implications for neural coding.}

\author{Thibaud Taillefumier\affil{1}{Laboratory of Mathematical Physics, Rockefeller University}
\and Marcelo O. Magnasco\affil{1}{}
}

\contributor{DRAFT: To Be Submitted to Proceedings of the National Academy of Sciences
of the United States of America}

\maketitle

\begin{article}
\begin{abstract}
Finding the first time a fluctuating quantity reaches a given boundary is a deceptively simple-looking problem of vast practical importance in physics, biology, chemistry, neuroscience, economics and industry. Problems in which the bound to be traversed is itself a fluctuating function of time include widely studied settings in neural coding, such as neuronal integrators with irregular inputs and internal noise. We show that the probability $p(t)$ that a Gauss-Markov process will first exceed the boundary at time $t$ suffers a phase transition as a function of the roughness of the boundary, as measured by its  H\"older exponent $H$, with critical value $H_c=1/2$. For smoother boundaries, $H > 1/2$, the probability density is a continuous function of time. For rougher boundaries, $H<1/2$, the probability is concentrated on a Cantor-like set of zero measure: the probability density becomes divergent, almost everywhere either zero or infinity. The critical point $H_c =1/2$ corresponds to a widely-studied case in the theory of neural coding, where the external input integrated by a model neuron is a white-noise process, such as uncorrelated but precisely balanced excitatory and inhibitory inputs. We argue this transition corresponds to a sharp boundary between rate codes, in which the neural firing probability varies smoothly, and temporal codes, in which the neuron fires at sharply-defined times regardless of the intensity of internal noise.
\end{abstract}


\abbreviations{FPT, first passage time; OUP,
Ornstein-Uhlenbeck process; LIF, leaky integrate and fire neuron}

\dropcap{A} Brownian process $W(t)$ which starts at $t=0$ from $W(t=0)\, =0$ will fluctuate up and down, eventually crossing the value $1$ infinitely many times: for any given realization of the process $W$ there will be infinitely many different values of $t$ for which $W(t)=1$. Finding the very first such time, 
$$\tau = \inf \{ t \, \vert \, W(t)=1\}$$ 
known as the {\em first passage} of the process through the boundary $L=1$, is easier said than done, one of those classical problems whose concise statements conceal their difficulty \cite{R1,R2,R3,R4}. For general fluctuating random processes the first passage time problem (FPTP) is both extremely difficult \cite{R5,R6,R7,R8,R9} and highly relevant, due to its manifold practical applications: it models phenomena as diverse as the onset of chemical reactions \cite{R10,R11,R12,R13,R14}, transitions of macromolecular assemblies \cite{R15,R16,R17,R18,R19}, time-to- failure of a device \cite{R20,R21,R22}, accumulation of evidence in neural decision-making circuits \cite{R23}, the ``gambler's ruin'' problem in game theory \cite{R24}, species extinction probabilities in ecology \cite{R25}, survival probabilities of patients and disease progression \cite{R26,R27,R28}, triggering of orders in the stock market \cite{R29,R30,R31}, and firing of neural action potentials \cite{R32,R33,R34,R35,R36,R37}.

Much attention has been devoted to two extensions of this basic problem. One is the first passage through a stationary boundary within a complex spatial geometry, such as diffusion in porous media or complex networks, as this describes foraging search patterns in ecology \cite{R38,R39}, and the speed at which a node can receive and relax information in a complex network \cite{R40,R41} .

The second extension is the first passage through a boundary that is a fluctuating function of time \cite{R42,R43,R44}, a problem with direct application to the modeling of neural encoding of information \cite{R45,R46}. This problem and its application are the subject of this paper. The connection arises as follows. The membrane voltage of a neuron fluctuates in response both to synaptic inputs as well as internal noise. As soon as a threshold voltage is exceeded, nonlinear avalanche processes are awakened which cause the neuron to generate an action potential or spike. Therefore the generation of an action potential by a neuron involves the first passage of the fluctuating membrane voltage through the threshold. This dynamics of spike generation underlies neural coding: neurons communicate information through their electrical spiking, and the functional relation between the information being encoded and the spikes is called a {\em neural code}. Two important classes of neural code are the {\em rate codes}, in which information is only encoded in the average number of spikes per unit of time (rate) without regard to their precise temporal pattern, and the {\em temporal codes}, in which the precise timing of action potentials, either absolute or relative to one another, conveys information.

Central to the distinction between rate and temporal codes is the notion of jitter or temporal reliability. This notion originates from repeating an input again and again and aligning the resulting spikes to the onset of the stimulus.  Time jittering is assessed  graphically through a raster plot and quantitatively in a temporal histogram (PSTH) which permits verifying the temporal accuracy with which the neuronal process repeats action potentials. A fundamental observation is that the very same neuron may lock onto fast features of a stimulus yet show great variability when presented with a featureless, smooth stimulus \cite{R33}. These two are extreme examples from a continuum---the jitter in spike times depends directly on the stimulus being presented \cite{R47} . 

\section{First passage through a rough boundary}

We shall make use of a simple geometrical construction, mapping the dynamics of a neuron with an input, internal noise and a constant threshold voltage, onto a neuron with internal noise and a fluctuating threshold voltage; the construction thus maps the input onto fluctuations of the threshold. We use as our model neuron the {\em leaky integrate-and-fire neuron} (LIF), a simple yet widely-used \cite{R36,R47,R48,R49,R50,R51,R52,R53,R54,R55,R56,R57,R58,R59} model of neuronal function defined by 
\begin{equation} \label{eq:Langevin}
\dot V = - \alpha V + I(t) + \xi(t)
\end{equation}
where $V$ is the membrane voltage, $1/\alpha$ is a decay time given by the $RC$ constant of the membrane, $I$ the current that the neuron receives as an input through synapses, and $\xi$ an internal noise. When $V$ first reaches a threshold value $l$ an action potential is generated, and the voltage is reset to zero. The nonlinearity of the model is concentrated on the spike generation and subsequent reset, so that between spikes we can integrate separately the effect of the input and of the noise: 
$$V=V_I+V_\xi  \qquad      \dot V_I 
 = -\alpha V_I + I(t) 
\qquad      \dot V_\xi = -\alpha V_\xi+\xi_t$$
Because the input $I(t)$ is fixed, the $V_I$ equation needs to be solved just once. Then the problem of $V(t)$ reaching the threshold $l$ can be recast as $V_\xi$ reaching the boundary $l-V_I $: we have transformed a problem with an input and a constant threshold into a problem with no input and a fluctuating threshold $l-V_I$. The reset operation $V=l \to V=0$ becomes $V_\xi=l-V_I\ \to\ V_\xi=-V_I$ (see Appendix).
 
These considerations lead us to examine the problem of the first passage time through a fluctuating threshold. In order to develop some intuition about the problem, we are going to break it up into two parts, a ``geometrical optics'' part, in which most first passages can be accounted for by simple ``visibility'' considerations, and a ``diffractive'' correction in which we take into account that random walkers can turn around corners. The geometrical part is simple: most first passages are generated by the walker running into a hard-to-avoid obstacle, as shown in Figure 1a. The intuition is that the walkers are moving left to right, rising onto a ceiling from which features are hanging, and as the walkers rise they collide with some feature. The problem is thus twice symmetry-broken: what matters are local minima of the boundary, not the maxima, which are hard to get into; and the walkers only spontaneously run onto the left flank of a local minimum. Therefore, a good first order approximation follows from observing that most of the first passages occur on the left flanks of local minima, and deeper local minima cast ``shadows'' on subsequent shallower minima.

\begin{figure}[b]
\centerline{\includegraphics[width=8.7cm]{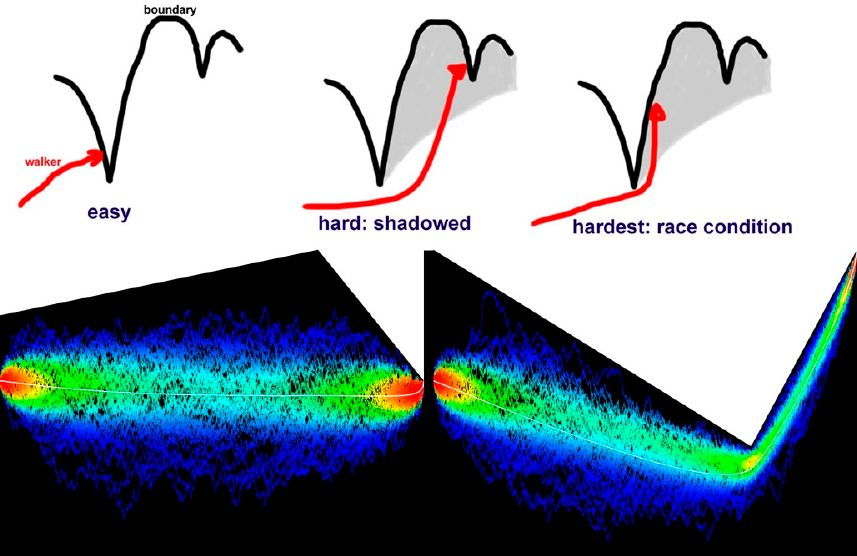}}
\caption{
How a random walk $V$ first hits a moving boundary $L$. In all panels, time $t$ is horizontal, the process $V$ and the boundary $L$ vertical. (A) It is highly probable to hit the left flank of a minimum, as the walkers are moving left to right and from the bottom up. (B) Each minimum ``casts a shadow'' behind it, so that hitting some features behind may be hard, as it requires missing the minimum, then rising sufficiently high to hit the second feature. (C) Hitting the right (rising) flank of a minimum is hardest, since it requires missing the minimum narrowly, then rising up, setting up a ``race condition'' between the boundary and the walker. Lower panels D and E: 300 sample paths which start at the red point on the left and have their first passage through the boundary (white) on the red point in the right. White curve: average trajectory (analytic). Sample paths are colored by the probability density of the point they go through. In (D), hitting a left flank of a minimum is easy, and the average trajectory to do so does not significantly deviate from the deterministic trajectory until the very end, where the white curve can be seen to rise onto the minimum following a square root. In (E), hitting the right flank of a minimum is hard, and the average trajectory to do so strongly deviates from the deterministic trajectories of the system, missing the minimum by just enough not to collide with it, then rapidly rising to meet the first passage point, again, in a square-root profile. 
}\label{fig1}
\end{figure}

\begin{figure}[bt]
\centerline{\includegraphics[width=7cm]{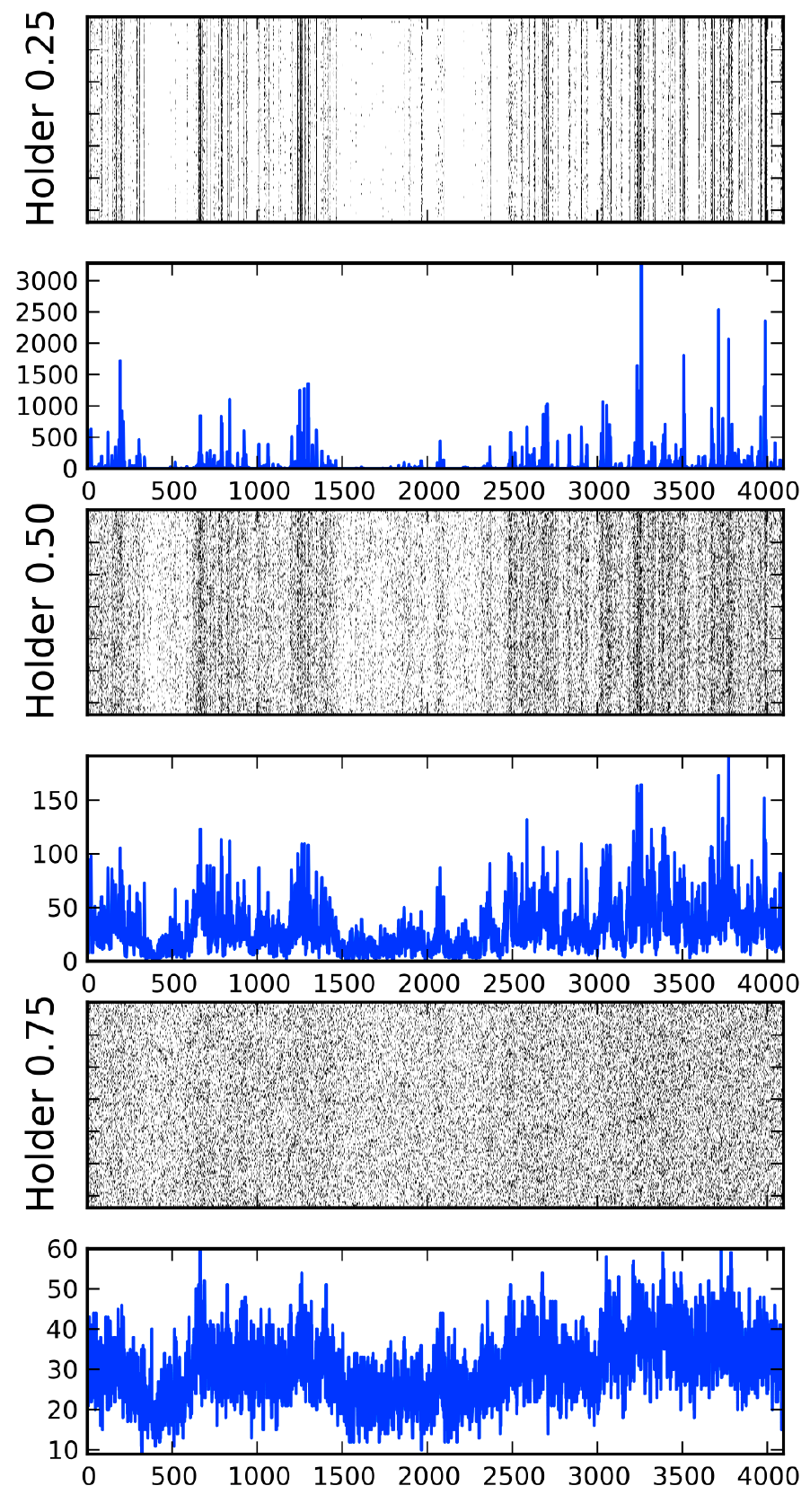}}
\caption{
Rasterplots and PSTH. A small segment of our dataset is displayed for clarity. A rasterplot and a plot of the PSTH are shown for each of three  H\"older exponents: 0.25 (rough), 0.5 (transition) and 0.75 (smoother, though still not differentiable). There's approximately the same number of spikes in all three groups. The rasterplots display the times at which the neuron fired (i.e. a first passage) stacked vertically (as a function of stimulus presentation number) to show repeatability. The PSTHs show a temporal histogram of said spikes. Please note the differences in vertical scale of the PSTHs: for H\"older exponent $H=0.75$ there are no bins with fewer counts than 10 events or more than 60, while for $H=0.25$ most bins have 0 counts while a few have over 1000 counts. }\label{fig2}
\end{figure}

\begin{figure}[bt]
\centerline{\includegraphics[width=8.7cm]{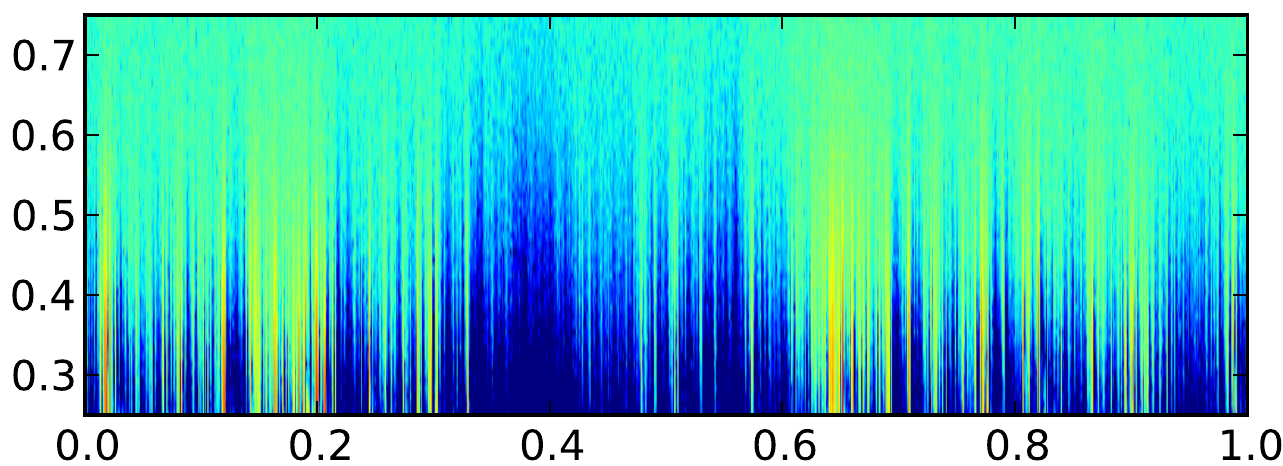}}
\centerline{\includegraphics[width=8.7cm]{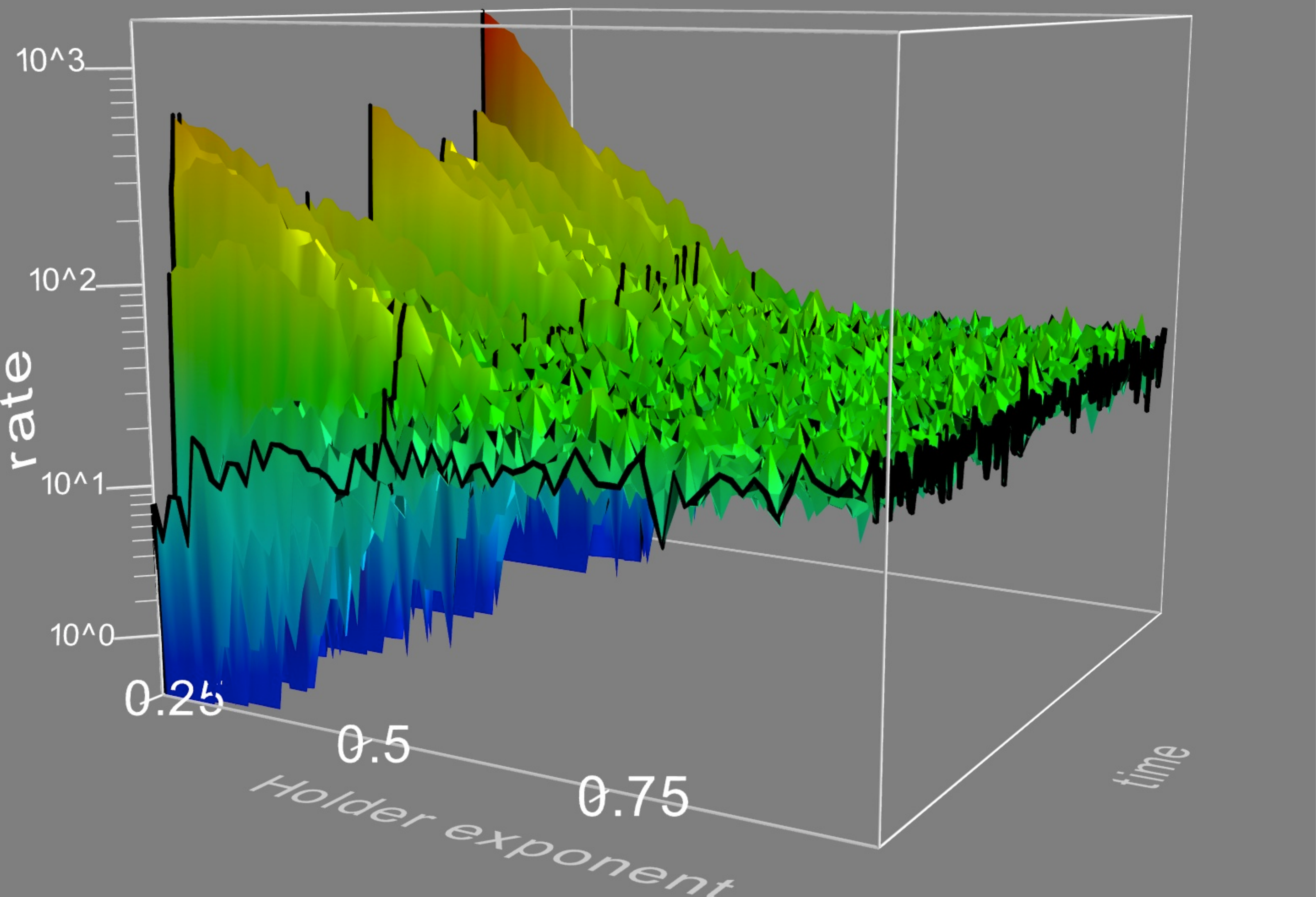}}
\caption{
(a) Probability density of firing as a function of time (horizontal) and  H\"older exponent (vertical), color coded in log scale. 51 values of the  H\"older exponent $H$ between 0.25 and 0.75 are stacked vertically. The bin counts shown in the PSTHs of Fig 2 are color coded with a logarithmic code. (b) 3D rendering of a section of the data in (a): vertical axis and color scale is logarithmic in the rate, where it is evident that towards the back of the figure ( H\"older exponent $H=0.25$) the rate either diverges or goes to zero a.e.
 }\label{fig3}
\end{figure}

However, there is a finite probability that a walker may narrowly avoid a local minimum and pass just under it, only to rapidly rise afterwards and hit the right rising flank of the barrier, as shown in Figure 1C. This is, effectively, a race between the boundary and the walker: if the walker can rise far faster than the boundary, then there is some probability of passage right of the minimum. But if the boundary rises faster than a walker can catch up with, then the probability of passage right of the minimum can be exponentially small. Let us consider a local minimum of the barrier $T(t)$ at time $t_0$ of the form 
$$T(t \ge t_0 ) \approx T(t_0) + (t-t_0)^h$$
and consider a walker that has just narrowly missed the minimum by an amount $\epsilon$: $W(t_0 )=T(t_0 )-\epsilon$. The probability of the process to be at value $W$ at time $t>t_0$ is, to leading order, 
$$P(W,t) \approx \exp\left({ -(W-W_0 )^2\over \Gamma(t-t_0) }\right)$$
and thus the probability of arriving at the barrier at time $t$ is approximately 
$$P_b (t)= \lim_{\epsilon\to 0} P(T(t),t) \approx \exp (-(t-t_0 )^{2H-1}/\Gamma)$$ 
When $H<1/2$ this expression has an essential singularity, its value singular-exponentially small for small times. 
In fact the probability and all of its derivatives are zero at $t_0$. For instance, consider a barrier whose flank to the right of the local minimum rises like $\sqrt[4]{\Delta t}$. As the fourth root in the barrier rises much more rapidly than the square root in the walker, the probability of hitting the barrier after the minimum looks like $\exp (-1/\sqrt{\Delta t})$, a function that has an essential singularity at $0$: the function as well as all of its derivatives approach $0$ as $\Delta t \to 0^+$.  

The parameter $H$ we described above, which is called the  H\"older exponent of the function, quantifies the ability of the barrier to, locally, rise faster or slower than a random walk. More formally, a function $f(t)$ is said to be $H$-H\"older continuous if it satisfies  
$|f(t)-f(s)|<C|t-s|^h$; the roughness exponent $H$ of the function is the largest possible value of $H$ for which the function satisfies a  H\"older condition. 
Up to now we have considered a single local minimum, and even though the probability of crossing is singular-exponential small for $H<1/2$, it is still nonzero. However, if the boundary is rugged, the local minima are dense. This density is not an issue for $H>1/2$, that is inputs which are smoother than the internal noise; in this case the probability density of first passages is nowhere zero. But when $H<1/2$ so the input is rougher or burstier than the internal noise, the probability density ceases to be a function: it is zero almost everywhere except for a set of zero measure where it diverges. 

\section{Results}

We postpone to the Appendices the more formal proofs of regularity  of the first passage time probability distributions. We proceed now, instead, to discuss  numerical simulations and their analysis. 
  
We carried out careful numerical integration of equation \eqref{eq:Langevin}, for all  H\"older exponents $H$ in the range $(0.25-0.99)$ in increments of $0.01$. In order for the results of the simulations at different  H\"older exponents to be directly comparable to one another, we generated the inputs $I(t)$ by using the exact same overall coefficients in the basis functions of the Ornstein-Uhlenbeck process described in \cite{R60}, but scaled differently according to the  
H\"older exponent laws (see Appendix). For each one of the $75$  H\"older exponents between $0.25$ and $0.99$, $62000$ repetitions of the stimulus were performed, accumulating $100.000.000$ first passages per  H\"older exponent. We computed the first passages using the fast algorithm described in \cite{R55,R60}, which carries out exact integration in intervals which are recursively subdivided when the probability that the process attains the first passage exceeds a threshold, in our case $10^{-20}$. The first passages were computed to an accuracy of $2^{-26}=1/67108864$, and the allowable probability that a computed passage is not in fact the first one is $p_{fail}= 10^{-15}$, so as to have an overall probability of $10^{-5}$ that any one of our 7.5 billion numbers is not in fact a true first passage. The values of the first passages were histogrammed in $2^{22}$ bins; this histogram, which we call our PSTH (peristimulus time histogram) in analogy to the term in use in neural coding, represents the instantaneous probability distribution of first passage integrated over the bins, or, equivalently, the finite differences over a grid of the cumulative probability distribution function for firing. 

\begin{figure}[bt]
\centerline{\includegraphics[width=6cm]{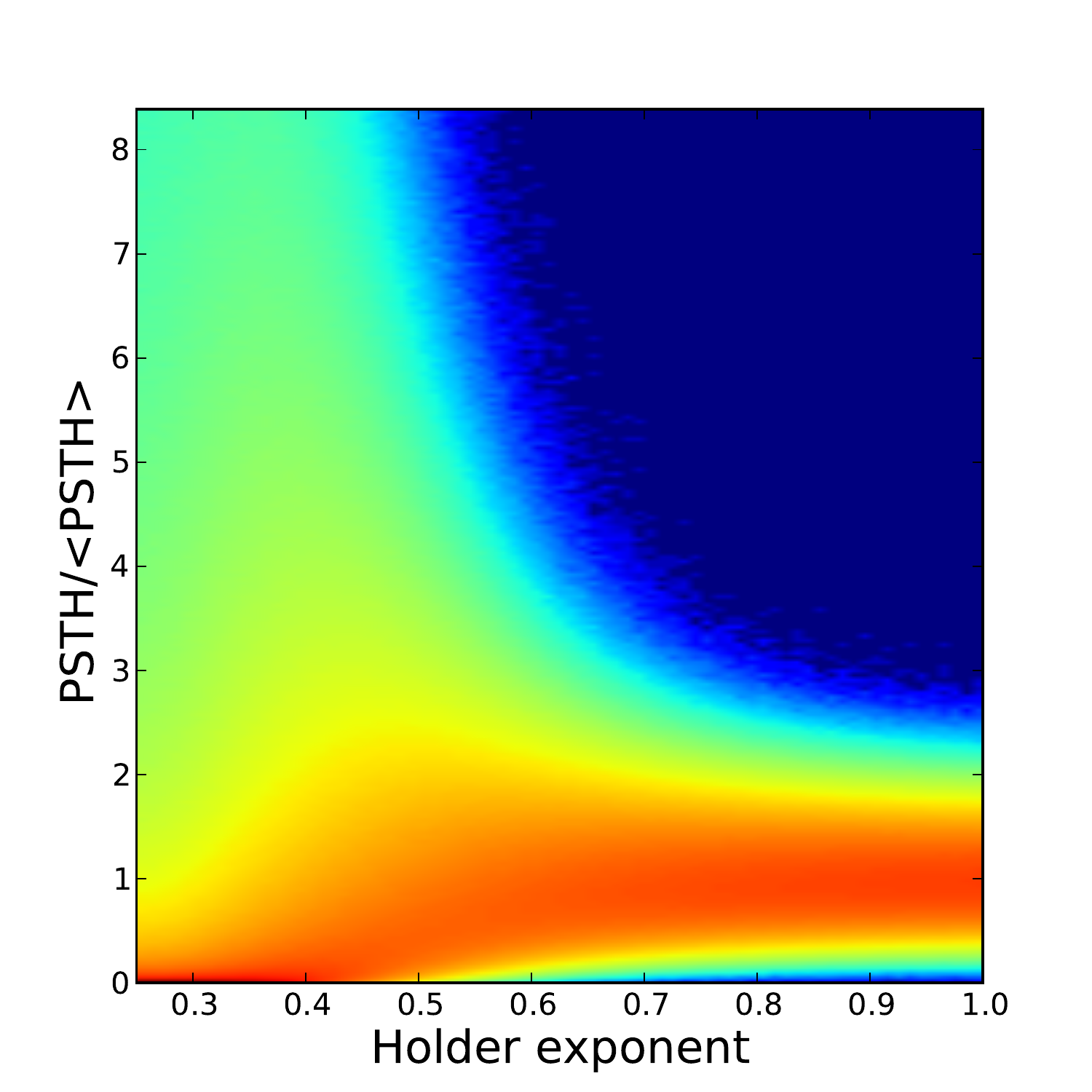}}
\centerline{\includegraphics[width=6cm]{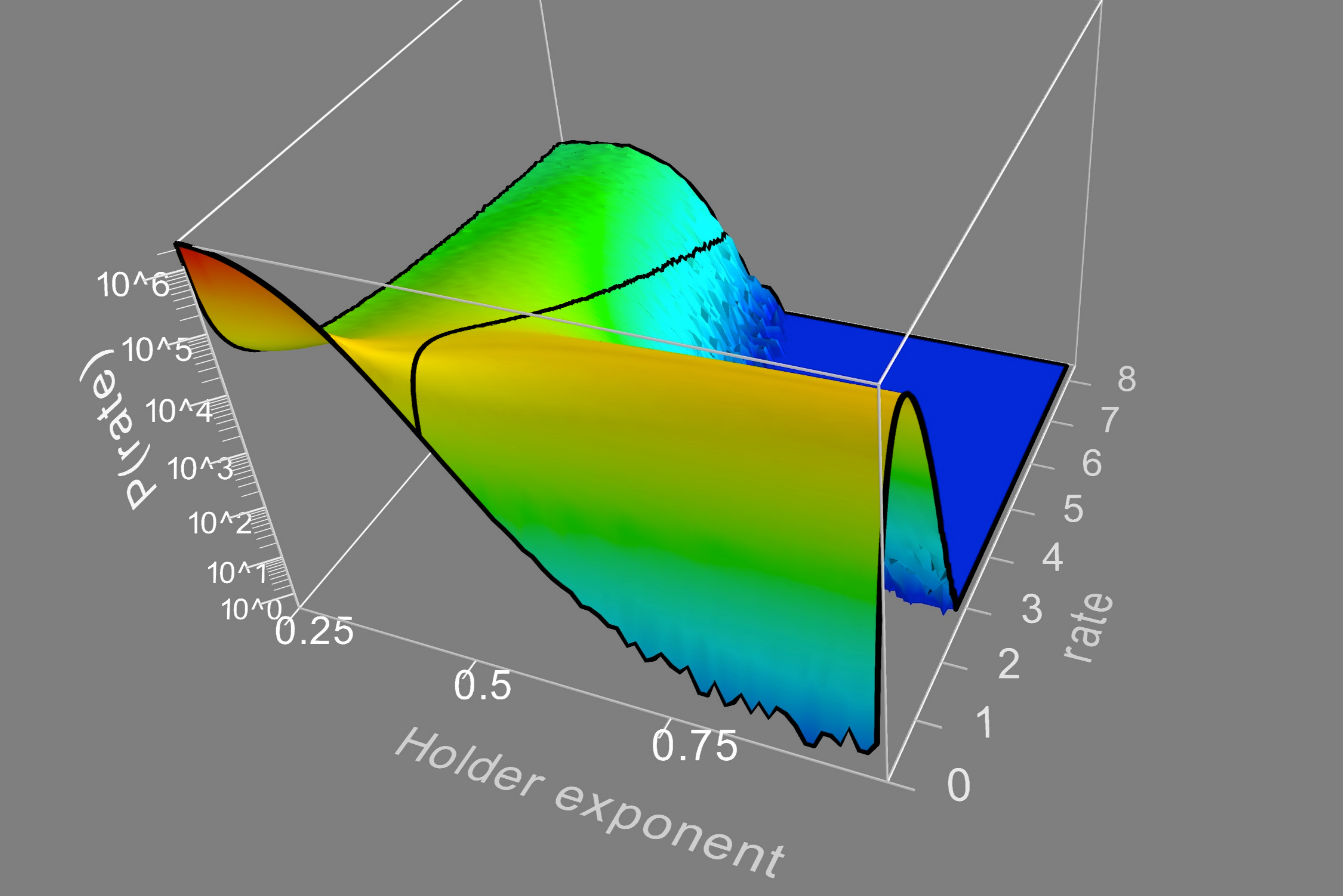}}
\caption{
Density map of PSTH bin counts. The individual bin counts of the PSTHs as shown in Figs 2 and 3 are histogrammed here, and the value displayed as a logarithmic color map. All 7.5 billion spikes in our dataset were used for this plot. The bin counts are normalized by the average bin count ($10^8 / 2^{22}$). For large  H\"older exponents, the probability of observing an actual count agrees with counting statistics given the average. As the  H\"older exponent becomes smaller, this distribution becomes wider, until below 0.5 it becomes heavy-tailed. Notice the bottom row of the figure, representing the probability of observing a bin with zero counts. It is zero for all $H>0.5$, becomes nonzero at $H=0.5$, and for $H<0.5$ it is the maximum of the distribution (i.e. the brightest red value).  }\label{fig4}
\end{figure}

The transition from smooth probability distribution to a singular measure is illustrated in Figures 2 and 3, where,  as the  H\"older exponent is lowered, the concentration of the first passage probability on a small set is evident. Histogramming the individual bins of the PSTH we get the probability distribution to observe a given instantaneous rate of firing, shown in Figure 4. For large  H\"older exponents the rate does not deviate far from its mean. However, as the  H\"older exponent becomes $1/2$, both the probability of observing a zero rate, as well as the probability of seeing a rate far larger than the mean, become substantial. For $H <1/2$ it becomes very probable to observe either zeros or large values of the instantaneous rate. 
This statement can be made precise by observing the tails of the probability distribution, and this is best accomplished, given our numerical setup, by looking at the tails of the cumulative probability distribution, namely 
$$F(x)=\int_{-°}^x P(x' )dx'$$
and then analyzing $1-F(x)$  vs $x$ for large $x$, which is carried out in Figure 5. Figure 5a shows that the tails of the distribution, when $x\gg 1$, decay exponentially for $H>1/2$ but behave like stretched exponentials when $H<1/2$:
\begin{eqnarray}
1-F(x)\approx &e^{-ax} \, ,&\qquad h>1/2 \, , \\
1-F(x)\approx &e^{-b\sqrt{x}} \, ,&\qquad h<1/2 \, .
\end{eqnarray}
This observation is quantified in Fig 5b, where $\log (1-F)$ is fitted with a quadratic polynomial in $\sqrt{x}$, namely 
$$-\log(1-F(x))\ \approx a x+b\sqrt{x}+c$$
For $H<1/2$ the quadratic coefficient in the fit, which gives the convergent linear term, vanishes, uncovering the stretched exponential behavior. This quantitatively proves our assertion of a phase transition at $H=1/2$.

\begin{figure*}[tb]
\centerline{\includegraphics[width=18cm]{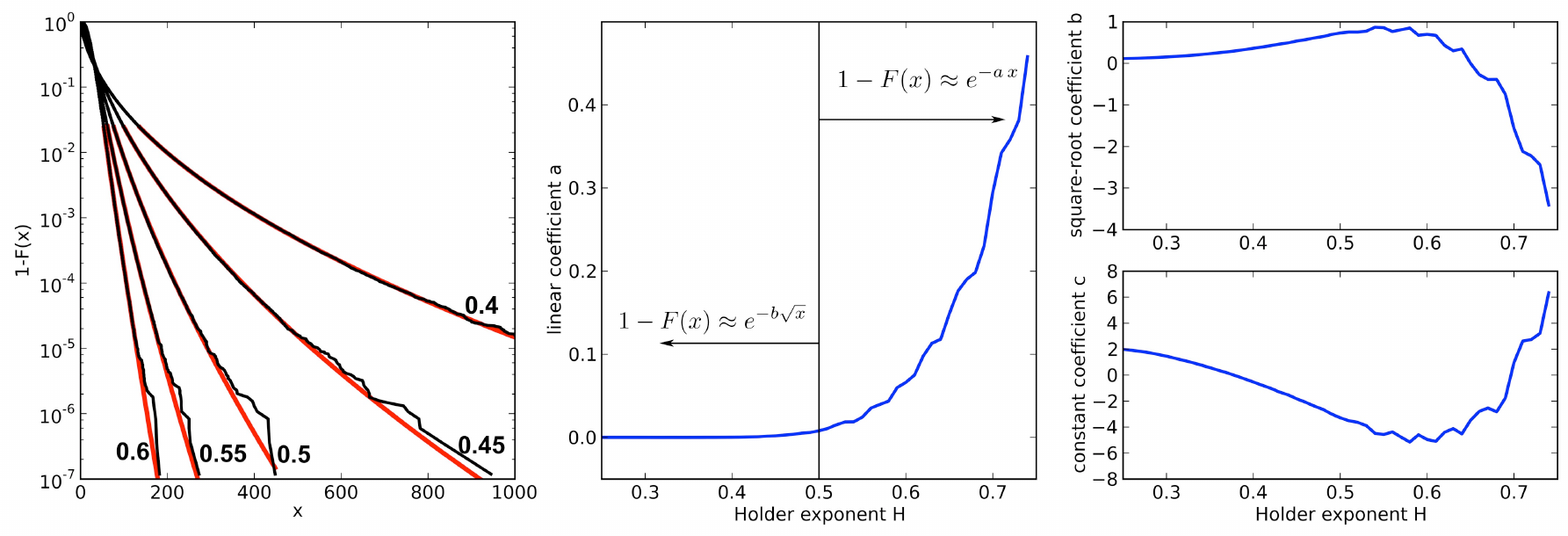}}
\caption{
The tail of the cumulative probability distribution of observing a given count in the PSTH becomes a stretched exponential at H\"older exponent $H=0.5$. Top, the tails of the cumulative probability distribution, plotted as $1-F(x)$ vs. $x$, for H\"older exponents $0.4,0.45,0.5, 0.55$ and $0.6$ (right to left). The probability distribution is minus the derivative of these curves. Superposed on the data (black) a fit to the last $10^5$ data points in the cumulative, i.e., the higher 2\% 
percentile (red), in the form $-\log(1-F(x))\ \approx a x+b\sqrt{x}+c$.  
Right, the coefficients $a$, $b$ and $c$ for the aforementioned fit, plotted as a function of the H\"older exponent $H$. Notice that the linear component a is (numerically) zero for $H<0.5$, exposing the $\sqrt{x}$  term as the next higher order. For $H>0.5$ the positive linear term guarantees convergence of all moments of the distribution. 
}\label{fig5}
\end{figure*}

\section{Discussion}

In abstract, mathematical terms, we have shown that the probability of observing a first-passage of a Gauss-Markov process through a rough boundary of H\"older exponent $H$ suffers a phase transition at $H=1/2$. 
The integral of the probability on equispaced grids becomes a stretched exponential, showing the underlying instantaneous probability has ceased to be a function:  it is concentrated on a Cantor-like set within which it is infinite, and it is zero outside this set.  
Gauss-Markov processes, such as the Ornstein-Uhlenbeck process, can be mapped to the canonical Wiener process through a deterministic joint scaling and time-change operation that preserves H\"older continuity\footnote{This transformation is referred as the Doob's transform.}.
Furthermore, being the solution to a linear Langevin equation, the first-passage problem for drifted Gauss-Markov processes can always be formulated in terms of a fluctuating effective barrier that integrates the drift contribution.
Therefore, our analysis directly applies to this situation.
As non-linear diffusions with bounded drift behave like Brownian motion at vanishingly small scales, we envision that our result is valid for this more general class of stochastic processes with H\"older continuous barrier.
However, in this case, the barrier under consideration does not summarize the drift contribution of the diffusion.

In terms of the original motivating problem, the encoding of an input into the timing of action potentials by a model neuron, this means that within our (theoretical and rather aseptic) model, there is an abrupt transition in character of the PSTH, the instantaneous firing rate constructed from histogramming repetitions of the same stimulus. The transition happens when the input has the roughness of white noise, conceptually the case in which the neuron is receiving a barrage of statistically independent excitatory and inhibitory inputs, each with a random, Poisson character. For inputs which are smoother than this, the PSTH is a well-behaved function whose finite resolution approximations converge nicely and properly to finite values. However, when the input is {\em rougher} than uncorrelated excitation and inhibition, for example when excitatory and inhibitory activities are clustered positively with themselves and negatively with one another, then the PSTH is concentrated on a singularly small set, which means that the PSTH consists of a large number of sharply-defined peaks of many different amplitudes, but each one of them having precisely zero width. The width of the peaks is zero {\em regardless} of the amplitude of the internal noise; increasing internal noise only leads to power from the tall peaks being transferred to lower peaks, but all peaks stay zero width. Since the set of peaks is dense, refining the bins over which the PSTH is histogrammed leads to divergencies. 

Concentration of the input into rougher temporal patterns would evidently be a function of the circuit organization. For example, in primary auditory cortex, the temporal precision observed in neuronal responses \cite{R61} appears to originate in the concentration of excitatory input into sharp ``bump''-like features \cite{R62}. 

It currently remains to be seen whether our mechanism will resist the multiple layers of real-world detail separating the abstract equation \eqref{eq:Langevin} from real neurons in a living brain. Obviously, the infinite-sharpness of our mathematical result shall not withstand many relevant perturbations, which will broaden our zero-width peaks into finite thickness. That this will happen is indeed sure, but not necessarily relevant, because a defining characteristic of phase transitions is that their presence affects the parameter space around them even under strong perturbations: that is why studying phase transitions in abstract, schematic models has been fruitful. Thus the real question remaining is whether our mechanism can retain  enough temporal accuracy to be relevant to understand the organization of high-temporal-accuracy systems such as the auditory pathways, and whether our description of the roughness of the input as the primary determinant of coding modality, temporal code or rate code, may illuminate and inform further studies.

\vskip 1cm
\appendix[Proofs]


Consider the stochastic leaky integrate-and-fire model for a spike triggering membrane threshold $l$ and a post-spiking reset value $r<l$.
Suppose a spike is emitted at time $t_i>0$.
With initial condition ${X_{t_i}^+} = r$, the inhomogeneous linear stochastic differential system 
\begin{equation}\label{eq:stochLin}
dX_t = -\alpha X_t \, dt + \sigma \, dW_t + dC(t) \, , \quad t>t_i  \, , 
\end{equation}
describes the ensuing sub-threshold noisy dynamic of the potential when driven by the input current $dC(t)$.
Here, $dC(t)$ shall be considered as the infinitesimal increment of a time-varying load function $C(t)$ that is $H$-continuous for a given H\"{o}lder exponent $H>0$, i.e. for every $T>0$, there exists a constant $c_T> 0$ such that for all $0<t,s<T$
\begin{equation*}
\lim_{\delta \to 0+} \sup_{ \vert t -s \vert \leq \delta} \frac{\vert C(t) - C(s)\vert}{{\vert t-s \vert}^h} \leq c_T \, .
\end{equation*}
Notice that, at the cost of rescaling $X$ and $I$ by $\sigma$, we can restrain ourselves to the study of the case $\sigma = 1$.


\section{Effective Barrier Formulation}

The nonlinearity of the leaky integrate-and-fire model lies entirely in the spike generation and subsequent reset, so that we can  separately integrate input and noise between spikes. 
Thus, our first-passage problem for constant threshold $l$ and varying forcing $dC$ becomes a first-passage problem without driving forces to a fluctuating effective barrier.
Precisely, we solve \eqref{eq:stochLin} writing $X = U^i+ l^i$, where we separate the stochastic part $U^i$ (the Ornstein-Uhlenbeck process obtained for $dC=0$) and the deterministic part $l^i$ arising from the integration of the input $dC(t)$:
\begin{eqnarray*}
U^i_t &=& r \, e^{-\alpha (t-t_i)} + \int^{t}_{t_{i}} e^{-\alpha (t-s)} \, dW_s \, , \\
l^i(t) &=&  \int^{t}_{t_{i}} e^{\alpha (t-s)} \, dC(s) \, .
\end{eqnarray*}
Determining the next spiking time $t_{i+1}$ can be cast in terms of a first-passage problem for the process $U^i$ with the effective barrier $t \mapsto L^i(t) = l - l^i(t)$:
\begin{equation}\label{eq:FPT1}
t_{i+1} = \inf \lbrace t>t_i \, \vert \, U^i_{t} > L^i(t)\rbrace \, .
\end{equation}
Therefore, a train of spikes $t_0 < t_1< \ldots <t_n$ is determined by solving consecutively the first-passage problems \eqref{eq:FPT1}.
Note that, due to the reset rule, the effective barriers do not agree at spiking times $L^{i-1}(t_{i}^-) \neq L^{i}(t_{i}^+) = l$. 
However, for all $i>0$, we have for $t>t_i$:
\begin{eqnarray*}
\Big \lbrace U^i_t < l-l^{i}(t) \Big \rbrace & = & \Big \lbrace  U^i_t - \int_{t_{i-1}}^{t_i}  e^{-\alpha (t-s)}dC(s)< l -l^{i-1}(t) \Big \rbrace \\
\end{eqnarray*}
Making  the left-hand term $U'^{i}_t$ of the second inequality explicit, we have
\begin{eqnarray*}
U'^i_t =  e^{-\alpha(t-t_i)} \left( r- \int_{t_{i-1}}^{t_i}  e^{-\alpha (t_i-s)}dC(s) \right ) +  \int^{t}_{t_{i}} e^{-\alpha (t-s)} \, dW_s  \, ,
\end{eqnarray*}
and we recognize $U'^i_t$ as the solution of \eqref{eq:stochLin} for $dC = 0$,  with the new initial condition:
\begin{equation*}
U'_{{t_i}^+} = r - \int_{t_{i-1}}^{t_i}  e^{-\alpha (t_i-s)}dC(s) = L^{i-1}(t) - (l-r) \, .
\end{equation*}
As a result, the train of spikes $t_0 < t_1< \ldots <t_n$ is determined by the sequence of first-passage problem:
\begin{equation}\label{eq:FPT2}
t_{i+1} = \inf \lbrace t>t_i \, \vert \, U'^{i}_{t} > L^{0}(t)  \rbrace \, ,
\end{equation}
where $U'^i$ is the standard Ornstein-Uhlenbeck process with initial condition $U'_{{t_i}^+} = L^0(t)- (l-r)$.
In other words, by altering the reset rule, the linearity of the stochastic dynamics allows us to recast the successive first-passage problems \eqref{eq:FPT1} in terms of a sequence of first-passage problems for one single continuous barrier $L=L^0$ \eqref{eq:FPT2}. \\


\section{First-Passage Markov Chain}

In a typical experiment, the spiking history of a neuron is recorded in response to repeated presentations of the same stimulus. 
We idealize this situation by studying the distribution of spiking events when an input cyclically forces a leaky-integrate-and-fire neuron. 
To avoid discontinuity effects, we choose a barrier satisfying $L(T) = L(0)$ for some $T>0$ and then extend the definition of $L$ on the whole time-line by periodization  $L(t)=L(t \;  \mathrm{mod} \;  T)$. 
Then, the sequence of random times $\mathcal{T}_n=(\tau_n  \: \mathrm{mod} \; T)$, where $\tau_n$ denotes successive first-passage times to $L$, defines a discrete-time Markov chain $\mathcal{T}$ over the finite time period $[0,T)$, seen as an oriented circle\footnote{ The passage of time orients the circle and we identify the future time $T$ with the past time 0}.\\ 
To make it more formal, assume we can choose a load function satisfying for some $T>0$
\begin{equation}\label{eq:contCrit}
\int_{0}^{T} e^{-\alpha (T-s)} \, dC(s) = 0 \, ,
\end{equation}
which amounts to having a periodic effective barrier by setting $L(t) = L(t \; \mathrm{mod} \; T )$.
For any time $s$ in $[0,T)$, consider the first passage time $\tau_s$ for an Ornstein-Uhlenbeck process starting at $U_{s} = L(t)- (l-r)$ and the barrier $L$.
Because $L(t)$ is a continuous function, it is known that the random variable $\tau_s$ admits a continuous non-decreasing cumulative distribution function $F_s:[s,\infty) \rightarrow [0,1]$ ~\cite{Lehm02}.
We then define the measure $k_s$ on the Borel sets of $[s,\infty)$ by setting for every open set $O_{a,b} = (a,b) \subset [s,\infty)$, $s<a<b$:
\begin{equation*}
k_s(O_{a,b}) = F_s(b) -F_s(a) \, . 
\end{equation*}
Moving forward, we identify $[0,T)$ with the circle $\mathbb{C}= \mathbb{R}/T\mathbb{Z}$, which is compact for the Euclidean distance and for which the open arc circles $O_{(a,b)}$, are oriented counter-clockwise from $a$ to $b$, and generate the collection of Borel sets $\mathcal{B}(\mathbb{C})$.
Equipped with the quotient map $\pi: \mathbb{R}^+ \sim \mathbb{C}$, we define on the compact measurable state space $\big(\mathbb{C},\mathcal{B}(\mathbb{C})\big)$ the measure kernels $k^T_s$ by setting for all open $O_{(a,b)}$
\begin{equation*}
k^T_s(O_{(a,b)}) = k_s\big(\pi^{-1}(O_{(a,b)})\big)  \, .
\end{equation*}
The collection of measures $k_s^T$ form  a transition kernel on the compact state space $\mathbb{C}$.
Given an initial probability measure $\mu_0$ on $\mathbb{C}$, they define a continuous state, discrete time Markov chain~\cite{Haggstrom:2002,Norris:1998,Stewart:2009} $\mathcal{T} = \left( \mathcal{T} , \mathcal{P} \right)$ on $(\Omega, \mathcal{M}) = \big(\mathbb{C},\mathcal{B}(\mathbb{C})\big)^{\mathbb{N}}$, whose probability $\mathcal{P}$ satisfies:
\begin{eqnarray*}
\lefteqn{
 \forall n \in \mathbb{N} \, , \quad \mathcal{P}(d\tau_n, \ldots , d\tau_0)= } \\
 && k_{\tau_{n-1}}(d\tau_n)\ldots k_{\tau_0}(d\tau_1) \mu_0(d\tau_0)\, .
\end{eqnarray*}
In particular, for all $u$, $v$ in $\mathbb{C}$, $v \mapsto k^T_s(O_{(u,v)})$ is continuous in $v$ with $k^T_s(\mathbb{C}) = 1$.\\
We shall see $k^T_s$ as the cumulative distribution of $\tau_n$ when the underlying process $U^n$ starts at $U_s^n=L(s) - (l-r)$, i.e. the distribution of a spiking event knowing that the previous spike occurs at $t$. 
As such, the kernels $k^T_s$ need not admit a density $\kappa$ satisfying $k^T_s(dt) = \kappa(s,t) \, dt$, similarly to the ``Devil's staircase'' resulting from the integration of the uniform measure over the triadic Cantor set~\cite{Mandelbrot:1982kb}.


\section{Ergodicity of the Markov Chain}\label{sec:Ergodicity}

We are interested in using this Markov framework to elucidate the distribution of spiking events when a neuron is driven cyclically by an input defined \eqref{eq:contCrit}.
To ensure that the instantaneous firing rate and the probability of spiking coincide, we show that the Markov Chain $(\mathcal{T},\mathcal{P})$ is \emph{ergodic}, a notion we define in the following.\\
An distribution $\mu$ is invariant by $(\mathcal{T},\mathcal{P})$ if it satisfies
\begin{equation*}
\mu(dt) = \int_{0}^{T} k^{T}_{s}(dt)\mu(ds) \, ,
\end{equation*}
so that if $\mathcal{T}_n$ is distributed according to $\mu$, so is $\mathcal{T}_{n+1}$. 
When there exists a unique such measure $\mu$, for any initial distribution $\mu_0$ and any measurable set $B$ on the circle $\mathbb{C}$
\begin{equation*}
\lim_{N \to \infty} \frac{1}{N} \sum_{n=0}^{N-1}  \mathbbm{1}_{B}(T_n) = \mu(B) \, , \quad \mathbbm{1}_{B}(x) =
\left\{
\begin{array}{ccc}
1  & \mathrm{if}  &  x \in B  \\
 0 &  \mathrm{if}  &   x \notin B
\end{array}
\right.
\, ,
\end{equation*}
and the Markov chain is said to be ergodic.
Simply stated, the mean sojourn-time of the Markov chain in $B$ tends toward the measure of $B$ under $\mu$.  \\
We can show that the Markov chain $(\mathcal{T},\mathcal{P})$ is indeed ergodic for $H$-continuous functions with $H>0$.
Since the state space $\mathbb{C}$ of $(\mathcal{T},\mathcal{P})$ is compact,  it is enough to show that it has the strong Feller property~\cite{Hernandez-Lerma:2003th} to prove the existence of invariant measures, i.e.
\begin{equation*}
\forall B \in \mathcal{B}(\mathbb{C}) \, , \quad s_n \rightarrow s \in \mathbb{C}, \quad \Rightarrow \quad k_{s_n}(B) \rightarrow k_{s}(B) \, .
\end{equation*}
To establish the unicity of the invariant measure $\mu$, it is enough to show that the Markov chain $(\mathcal{T},\mathcal{P})$ has the irreducible property~\cite{Hernandez-Lerma:2003th}:
\begin{equation*}
\forall B \in \mathcal{B}(\mathbb{C}) \, , \quad \forall s \in \mathbb{C} \, , \quad k_{s}(B) > 0 \, .
\end{equation*}
We deduce the two properties above from consideration about  the first-passage time problem in Supplementary Materials.\\
The Feller property specifies that, if two identical leaky integrate-and-fire neurons spike respectively at times $s$ and $t$, then, when $s$ asymptotically approaches $t$, the probability that the first neuron later spikes in a given time interval becomes the same as for the other neuron.
In other words, close initial conditions entail similar probability  laws for the occurrence of the next spiking events (in the sense of the Kolmogorov test).\\
The irreducible property, which states that if one spiking time is achievable for a given starting condition (previous reset time), it is attainable for any starting time, similarly stems from these two intuitive observations.
If one trajectory starting at $t$ has a non-zero probability to hit this barrier in a given time region, we can easily convince ourselves that another trajectory starting at any $s$ has a non-zero probability to be close to the reset value in $t$, and from there, unfold as a trajectory that has been reset in $t$.\\
Intuitively, these properties  holds for our first-passage Markov chain for two reasons.
First, the continuity of the barrier which ensures the continuity of the cumulative distributions  of the transition kernels.
Second, the non-zero reset rules which constrain the membrane potential to be reset away from the barrier, thus avoiding pathological situations such as immediate absorption.


\section{Numerical Simulation of the Markov Chain}\label{sec:NumMethMarkovChain}

If the first-passage Markov chain $(\mathcal{T},\mathcal{P})$ is ergodic, due to the possible irregularity of the barrier, numerical simulation of its invariant measure demands that we resort to an approximation scheme. 
To justify this approach, we adapt a general result from~\cite{Karr:1975}, clarifying in which sense a sequence of Markov chains $(\mathcal{T}^N,\mathcal{P}^N)$ converge toward a limit chain $(\mathcal{T},\mathcal{P})$ when $N$ tends to infinity.\\

\noindent {\bf Theorem 2} (adapted from~\cite{Karr:1975}):
{\it Let $(\mathcal{X}^N,\mathcal{Q}^N)$ be a sequence of strongly Feller Markov chains defined on a compact state space $\mathbb{S}$.
If, 
for any $s$ in $\mathbb{S}$, the kernel probability measures $q^N_s$ of $\mathcal{X}^N$ converge in law toward a limit probability measure $q_s$,
then,
any limit in law of a sequence $\nu_n$ of invariant measures  of $(\mathcal{X}^N,\mathcal{Q}^N)$, is an invariant measure of the Markov chain $(\mathcal{X},\mathcal{Q})$ corresponding to the limit kernel $q$.}\\

In particular, if all $(\mathcal{X}^N,\mathcal{Q}^N)$ and $(\mathcal{X},\mathcal{Q})$ are ergodic, the sequence $\nu_n$ is uniquely defined and so is its limit distribution $\nu$, which is the stationary measure of $(\mathcal{X},\mathcal{Q})$.\\
For our purpose, an efficient approximation strategy of $\mu$ consists in exhibiting a sequence of ergodic strongly Feller Markov chains $(\mathcal{T}^N,\mathcal{P}^N)$ whose kernels $k^{N}_s$ converge to $k^T_s$ in law. 
This is accomplished by considering a sequence of first-passage Markov chains $(\mathcal{T}^N,\mathcal{P}^N)$ defined for the piecewise continuous periodic barriers $L_N$ that interpolates $L$ on the dyadic points $D_N = \lbrace k2^{-N}T \, \vert \, 0 \leq k < 2^{N}\rbrace$:
\begin{equation*}
L_{N}: t \in \mathbb{C} \mapsto \Exp{ U_t \, \vert \, U_{k2^{-N}T} = L(k2^{-N}T) \, , \: 0 \leq k < 2^{N} } 
\end{equation*}
where $\mathbb{E}$ denotes the expectation with respect to the law of $U$  (see~\cite{Taillefumier:2010jl}). 
Such Markov chains are ergodic by the same argument as for $(\mathcal{T},\mathcal{P})$.
Moreover, since we restrain ourselves to barriers $L$ that are $H$-continuous, the sequence $L_{N}$ converges uniformly toward $L$ (see Supplementary Materials), which in turn, implies the convergence in law (and in distribution) of $k^{N}_s$ toward $k^{T}_s$.
This demonstrates the cogency of approximating $L$ by $L^{N}$.  \\


\section{Frozen Noise as Injected Current}\label{sec:FrozenNoise}

In addition to providing a valid numerical method, the previous approach provides an easy description of the input $dC$ that gives rise to $L$.
The central results is adapted from~\cite{Thibaud:2011uq}:\\

\noindent {\bf Theorem}:
There exists a  Schauder basis of continuous functions $\psi_{n,k}$ compactly supported on $S_{n,k} = [k2^{-n+1}T, (k+1)2^{-n+1}T]$ such that, for all $N>0$,
\begin{eqnarray*}
\Exp{ U_t \, \vert \, U_{k2^{-N}T} \, , \: 0 \leq k < 2^{N} }  = \sum_{0 \leq n<N} \sum_{0 \leq k < 2^{n-1}} \psi_{n,k}(t) \cdot \Xi_{n,k}
\end{eqnarray*}
where the $\xi_{n,k}$ are the independent standard Gaussian variables
\begin{eqnarray*}
\Xi_{n,k} = \int_{0}^{T} \phi_{n,k}(t) \, dW_t \, , \quad \phi_{n,k} = \psi'_{n,k} - \alpha \psi_{n,k} \, .
\end{eqnarray*}
and the thus-defined functions $\phi_{n,k}$ form an orthonormal system of $L^2[0,T]$.\\

Equipped with this result, it is easy to see that writing the input $dC$ as a ``Gaussian white noise''
\begin{equation*}
dC_t = \sum_{0 \leq n} \sum_{0 \leq k < 2^{n-1}} \phi_{n,k}(t) \cdot \Xi_{n,k} \, , \quad \Xi_{n,k} \quad \, i.i.d \sim \mathcal{N}(0,1) \, ,
\end{equation*}
the statistics of the resulting random barrier 
\begin{equation*}
L_t = l- \int_{0}^{t} e^{-\alpha(t-s)} dC(s) = l- \sum_{0 \leq n} \sum_{0 \leq k < 2^{n-1}}  \psi_{n,k}(t) \cdot \Xi_{n,k} \, , 
\end{equation*}
is the same as for an Ornstein-Uhlenbeck process centered  around zero and translated upward by $l$.
Moreover, setting $\xi_{0,0} = 0$, we naturally enforce the periodic condition $L(t) = L(T) = l$. \\
However, we aim at studying the distribution of spiking events of a neuron cyclically driven by a deterministic input.
Accordingly, suppose now $dC(t) = dC_t(\omega)$ is a realization of our ``Gaussian white noise'', i.e. a frozen noise.
Then, $L(t) = L_t(\omega)$ is the sample path of an Ornstein-Uhlenbeck bridge translated upward, which is almost surely $H$-continuous  of exponent ${1/2}$.
For this reason, we denote such an input $dC^{1/2}$, the associated barrier $L^{1/2}$ and  the coefficients $\xi^{1/2}_{n,k}$.


\section{Family of H\"older Continuous Barriers}\label{sec:HolderBarrier}

From there, let us consider $\Omega_\xi$ the set of coefficients $\xi_{n,k}$ for which the continuous barriers of the form
\begin{equation*}
L_{N}(t) = l- \sum_{0 \leq n<N} \sum_{0 \leq k < 2^{n-1}}  \psi_{n,k}(t) \cdot \xi_{n,k} \, , 
\end{equation*}
converge uniformly on $\mathbb{C}$.
It can be shown~\cite{Thibaud:2011uq} that $\Omega_\xi$ contains the set 
\begin{equation*}
\Omega'_\xi = \lbrace \xi_{n,k} \in \mathbb{R}^{\mathbb{N}}  \, \vert \,  \exists \; \delta <1,  \exists \; N>0,  \forall n > N, \max_{k} \vert \xi_{n,k}\vert \leq 2^{n\delta/2} \rbrace \, .
\end{equation*}
From this, we deduce that  given $L^{1/2}$, for any real $H$ such that $0 < H<1$, the barrier $L^H$
\begin{equation*}
L^H(t) = l- \sum_{0 \leq n} \sum_{0 \leq k < 2^{n-1}}  \psi_{n,k}(t) \cdot  \,\xi^H_{n,k} \, , \quad\xi^H_{n,k} = 2^{n(H-1/2)} \, \xi_{n,k} \, ,
\end{equation*}
is well-defined as a continuous function of $\mathbb{C}$.
Keeping this in mind, we have at our disposal a well-known result~\cite{Meyer:1998} relating the local H\"older exponent of a function to the asymptotic behavior of the coefficients of its decomposition in the Schauder basis.
Adjusting to our situation, it directly entails that for all $H$, $0<H<1$, the barriers $L^H$ are almost-surely $H$-continuous.
Therefore, we can continuously (in the $L^\infty$-norm) control the asymptotic H\"{o}lder continuity of the effective barrier driving the activity of a leaky integrate-and-fire neuron by smoothly changing the coefficient $\xi^H_{n,k}$ used to construct piecewise approximations $L^H_N$. \\
In order to emphasize the effect of the varying H\"{o}lder regularity, we adopt a slightly modified version of our barriers $L^H$, by weighting them with a continuous function $H \mapsto c(H)$ under the from $L'^H=c(H) \big(L^H-L^H(0)\big)+L^H(0)$.
The function $c$ is chosen so that the newly formed barriers cause the neuron to fire with an overall mean firing rate (as opposed to the instantaneous mean firing rate which is time-dependent) remains constant when changing $H$.
Formally,  this constraint is equivalent to holding a constant mean inter-spike time 
\begin{eqnarray*}
\int_0^T \left( \int_s^{\infty} (t-s) \kappa^H_s(dt) \right) \, \mu_H(ds)
\end{eqnarray*}
while varying $H$\footnote{Notice that for the sake of well-posedness, the kernels that intervene in the formulation of the mean inter-spike time are computed for a periodic barrier $L^H$ but defined on $[0,\infty)$ instead of being wrapped on $[0,T)$.}.


\section{Integral Equation for the First-Passage Time}

We  establish the existence of a density function for the first-passage time of a Wiener process hitting a $H$-continuous barrier with $H>1/2$.
This property is formally referred to as the absolute continuity of the first-passage time distribution with respect to the Lebesgue measure on the real half-line.
Without loss of generality, we adopt the point of view of a killed Wiener process absorbed on a fluctuating boundary, which allows us to use the powerful machinery of the heat equation.
The presented result stems from the ground-breaking work of Gevrey~\cite{Gevrey:1913fk}  about parabolic differential equations, later actualized in a modern form by Rozier~\cite{Roz84}.\\
Integral equations for  the cumulative distribution of the first-passage time of a Wiener process naturally arise from probabilistic arguments.
Consider the event $\lbrace W_t > x\rbrace$ for an continuous barrier $L$ satisfying  $x>L(t)$. 
Then, the first-passage time $\tau$ with $L$ occurs certainly before $t$ and we can condition this event with respect to $\tau$, which yields 
\begin{equation} \label{eq:ProbVolt}
\mathbb{P}(W_t > x) = \Exp{\mathbb{P}(W_t > x \, \vert \, \tau)} = \int_0^{t}  \mathbb{P}(W_t>x \, \vert \, \tau = s) q(ds) \, ,
\end{equation}
where $q$ denotes the first-passage time probability measure.
Using the strong Markov property, on $\lbrace \tau=s \rbrace$, we can disregard the past-trajectory of $W$ and equate the probabilities $\mathbb{P}(W_t>x \, \vert \, \tau = s)$ and $\mathbb{P}(W_{t-s}>x-L(s) )$.
Differentiating equation \eqref{eq:ProbVolt} with respect to $x$, we end up with
\begin{eqnarray*}
 k\left(\frac{x}{\sqrt{t}} \right) = \int_{0}^{t} k{\left( \frac{x-L(s)}{\sqrt{t-s}}\right)}  q(s) \, ds \, ,
\end{eqnarray*}
where $k$ denotes the Heat kernel.\\
It is important to observe that as long as $L$ is $H$-continuous with $H>1/2$, we have
\begin{equation*}
\lim_{\tau \to t^-} \frac{L(t)-L(\tau)}{\sqrt{t-\tau}} = 0 \, .
\end{equation*}
Since $k$ is a smooth function, we can make the arbitrary value $x$ tend toward the barrier $L(t)$ by superior value and, through the dominated convergence theorem, we get the following integral Volterra equation~\cite{Park:1974,Park76}:
\begin{eqnarray}\label{eq:firstHeatVolterra}
k \left(\frac{L(t)}{\sqrt{t}} \right) = \int_{0}^{t} k{\left( \frac{L(t) - L(s)}{\sqrt{t-s}}\right)}  q(s) \, ds \, .
\end{eqnarray}
This integral equation, which dates back original work from Siegert~\cite{Siegert},  stems from the fact that $k(s,x;W_s,t)$ indexed by $s$ is a martingale~\cite{Strook:2006}, which offers a convenient way to generalize this equation to general time-inhomogeneous diffusion processes.


\section{Absolute Continuity of the First-Passage Time}

The integral equation is of the Volterra type, which comes in two flavor: equations of the \emph{first kind} and of the \emph{second kind}~\cite{Linz:1985}.
To ensure the existence and unicity of a solution to the equations of the second kind, we have the following powerful result:\\

\noindent {\bf Theorem} (adapted from~\cite{Hilbert:1962,Tricomi:1985}):
The linear Volterra equation of the second-kind
\begin{eqnarray*}
g(t) = f(t) + \int_0^t K(t,s) f(s) \, ds \, ,
\end{eqnarray*}
where $g$ is a piecewise continuous function has a unique piecewise continuous solution $f$ for all $t>0$ if $K$ is bounded on $0<s<t$  and if there exists a monotone increasing function $a$ with $\lim_{t \to 0} a(t) = 0$, such that for all $0<s<t$
\begin{equation*}
\int_s^t \vert K(t,\tau) \vert \, d\tau \leq a(t-s) \, .
\end{equation*}

Unfortunately, equation \eqref{eq:firstHeatVolterra} is a Volterra equation of the first-kind and as such cannot be dealt with directly.
However for barriers $L$ that are $H$-continuous, it can be recognized as a linear generalized Abel integral equation, that is an equation of the type
\begin{eqnarray*}
g(t) = \int_s^t \frac{K(t,\tau)f(\tau)}{{(t-\tau)}^h} \, d\tau
\end{eqnarray*}
where $f$ is the unknown, $g$ is a continuous function, and $K$ is a continuous kernel for $s \leq t$ and $0<h<1$.\\
Abel integral equations are frequently encountered in physics and there are methods to prove the existence and unicity of a solution by transforming the original equation into a Volterra equation of the second-kind.
In our case, it proceeds through the use of the Abel integral transform, which is designed to solve the canonical Abel equation
\begin{eqnarray*}
g(t) = \int_s^t \frac{f(\tau)}{\sqrt{t-\tau}} \, d\tau \, .
\end{eqnarray*}
The unique solution is given as
\begin{eqnarray*}
f(t) = \mathcal{A}[g](t) = \frac{1}{\pi} \frac{d}{dt} \left[ \int_s^t \frac{g(\tau)}{\sqrt{t-\tau}} \, d\tau \right]
\end{eqnarray*}
where $\mathcal{A}$ is the Abel inverse operator.
The application of $\mathcal{A}$ to equation \eqref{eq:firstHeatVolterra} reduces the problem to a Volterra equation of the second-kind: \\

\noindent {\bf Proposition} (adapted from~\cite{Roz84}):
If $L$ is $H$-continuous with $H>1/2$, through the application of the Abel operator, the Volterra equation of the first-kind \eqref{eq:firstHeatVolterra} is equivalent to the Volterra equation of the second-kind
\begin{eqnarray*}
\sqrt{2\pi}\mathcal{A}[g](t) =  q(t) + \frac{1}{\pi} \int_s^t K(t,\tau) q(\tau) \, d\tau \, ,
\end{eqnarray*}
with the kernel $K$  being defined as
\begin{eqnarray*}\label{eq:abelVoltDirich}
K(t,\tau) = \frac{\partial}{\partial t} \left\{ \displaystyle{\int_\tau^t} \frac{e^{ -\frac{\big(L(\sigma)-L(\tau)\big)^2}{2(\sigma-\tau)} }}{\sqrt{(t-\sigma)(\sigma-\tau)}}  \, d\tau \right\}  \, ,
\end{eqnarray*}
and $g$ denotes the continuous function $g(t) = k\big(s,x;t,L(t)\big)$.\\

A careful study shows that the kernel $K$ satisfies the conditions of Theorem ~\cite{Roz84}.
Thus the integral equation \eqref{eq:abelVoltDirich} obtained through the Abel transform admits a unique continuous solution, which is the density of the first-passage time to the barrier $L$.

\begin{acknowledgments}
This work was partially supported by NSF under grant EF-0928723.
We are indebted to Jonathan Touboul for helpful comments. 
\end{acknowledgments}



\bibliographystyle{plain}


\end{article}
\end{document}